# The relationship between professional and general ethics in generative AI

**Omri Asscher**
Department of English
Literature and Linguistics
Bar-Ilan University
omri.asscher@biu.ac.il

**Mirco Musolesi**
Centre for Artificial Intelligence
Department of Computer Science
University College London
Department of Computer Science
and Engineering
University of Bologna
m.musolesi@ucl.ac.uk

**Abstract**

Recent years have seen a growing discrepancy in the field of AI alignment: research and policy recommendations on AI ethics tend to assume a *general* set of ethical values, yet proliferating practice-specific uses of AI systems on the ground - in the legal, medical and translation domains, among others - have been effectively manifesting ethics of *professional practice*. This article begins by outlining the reasons why general and professional ethics are increasingly conflicted in contemporary AI systems, and by surveying how the research literature attests to, but has not yet resolved, this conceptual and practical challenge. We then conceptualize the main dimensions of AI models' decision-making in areas of professional practice, emphasizing professional ethics' *hierarchically structured relationship* with general ethics, and elaborating on the mechanisms through which they reach an equilibrium in situational contexts that involve conflict. It is through this equilibrium, we suggest, that certain professional ethics are prioritized over others and implemented in practice. We then show how our framework can be the basis for a systematic empirical assessment of AI models' professional ethics in various domains, identifying the nuances of the models' favored ethic by examining their production in a series of similar but not identical scenarios. Finally, we propose a formulation for how to intervene in and change AI models' favored ethics in professional practices – while noting the inherent dimension of subjectivity involved in both the evaluation and implementation of professional ethics in AI models.

**Key words**

professional ethics, general ethics, generative AI, AI alignment, reflective equilibrium, deontological threshold, AI ethical agency, ethical evaluation, large language models

## 1. Introduction

Recent years have seen a growing discrepancy between how AI ethics are discussed in the research literature and public discourse, on the one hand, and how AI decision-making has taken place in ethically-implicated cases of professional practice, on the other. Most academic and international protocols published about AI ethics, and studies in the field of AI alignment, tend to assume a *general*, even universal, set of ethical values (Hagendorff 2024). On the ground, however, practice- and profession-specific uses of AI systems have been occurring on a global scale, effectively manifesting an ethics of *professional practice*. In many cases, the general and professional ethical repertoires are in some tension with each other, creating a conceptual and practical challenge for AI ethics, with increasing implications. This is evident in AI production in areas and practices such as translation, medical advice, computer programming, or legal work - whether initiated and consumed by lay users or generated autonomously by AI systems and then consumed by lay users (Chien and Kim 2024, Mahajan 2025). The aims of the current article are to draw attention to this growing, yet still understudied, phenomenon, to frame it conceptually, and to outline the practical and empirical implications of this framework.

The reasons why this challenge has arrived at our doorstep at this particular time are revealing in themselves. The most foundational reason is the holistic, multitask nature of generative AI systems. This condition may seem self-evident, but it is in fact relatively new to AI ethics, coming in the wake of contemporary LLMs in the early 2020s. Today's AI systems are human-like in that their diverse decision-making, done in various contexts and related to different forms of expertise, is done by the *same entity*. Just like human actors, then, the selfsame AI entity makes decisions that involve a *particular expertise* (for example, choosing how to translate a rhetorical phrase in a text, how to code a program, or how to convey medical advice) while also, at the same time, having *general* traits that are not specific to this particular expertise. These general traits include cognitive features (e.g., how AI processes and produces language), socio-ideological features (e.g., the power relations and conventions the model reflects), cultural tendencies, and also, most important for our purposes, some form or other of general ethics.

The implications of the holistic nature of generative AI for our topic can be usefully understood by reference to the seminal debates on "narrow" AI and "general" AI (Pennachin and Goertzel 2007). In the major conceptual attempts to capture the difference between AI systems whose abilities encompass solving a wide range of problems across various domains, and AI systems whose ability is to do so only in a specific, delimited domain, today's conversational LLMs would be considered as a classic case of general AI (cf.

Kurzweil 2014). Yet, for our purposes, it is worth clarifying that LLMs are, in practice, *both* general AI and narrow AI at the same time. They do not have two separate independent decision-making entities, or "minds" – or, at least, no more than humans do (Brown et al. 2020, Reed et al. 2022). This, both general and narrow character of generative AI models, means that our conceptualization of their ethical agency needs to address how general and professional ethical repertoires *intersect with* and *inform* each other - their relationship effectively shaping the ultimate outcome of generative AI systems' practice-specific actions from an ethical perspective.

Beyond the holistic nature of today's LLMs, there are also some socio-technical reasons for the increased scope and relevance of the tension described here between general and practice-specific AI ethical agency. The growing customization and personalization of LLMs has encouraged the individual use of LLMs for professional practices (Salemi et al. 2024), which has led to the vastly growing use of AI for practices and expertise that had once involved mainly human professionals (Mahajan 2025), and also to an expansion of these practice-specific activities into areas that did not involve human professionals to begin with (e.g., using AI for translation in cases in which one would not otherwise hire a human professional translator but rather forgo the activity altogether, cf. Vieira et al. 2023; for our purposes, the term 'practice-specific ethics' is used analogously to 'professional ethics'). The growing autonomy of AI agents, especially as they become commercially widespread, has intensified this trend even further (Wang et al. 2025). These reasons, separately and jointly, have led to more and more AI decision-making being done, by LLMs serving the needs of lay users, at the intersection between **professional ethics** and **general ethics**.

The remainder of this article is structured as follows. The next section aims to describe the relevant research literature on professional and AI ethics, exemplifying the gap in the literature we wish to fill, and the themes and assumptions in the ethics literature we draw on. Then, the article conceptualizes the four main dimensions of the understudied phenomenon of practice-specific AI ethics. Conceptualizing these four dimensions is not merely a goal unto itself; rather, it can serve, as we conclude by suggesting, to ground a systematic *empirical* examination of AI professional ethics in practice, and a way to intervene in their orientation and decision-making.

## 2. AI ethics and professional ethics: main themes in the research literature

While much attention has been given to the effects of AI on employment (Deranty and Corbin 2024), and the consequences of AI use in professional settings such as legal workflows (Cardoso 2026), the research literature and policy initiatives on AI ethics and moral alignment have paid relatively little attention, so far, to how practice-specific ethics are embedded in generative AI decision-making and production. This is true for deontological, consequentialist and virtue-based ethical orientations, as well as hybrid configurations (Bench-Capon 2020, Krah and Tröschel 2026). Thinkers and scholars in the field tend to assume the existence of a *holistic*, coherent system of values, and imply that universal human ethics and rights, rooted in universal concerns, are sufficient to inform the conceptual and practical framework of AI ethics – including with regard to professional practices (Gabriel 2020, Unver 2026). Discussions that do admit ethical pluralism focus on issues of value pluralism or cultural relativism and remain on a largely philosophical level (Gabriel et al. 2025). They may debate the cultural "universality" of the morals AI is expected to align with, yet once they advance alternative paradigms of ethics (such as non-western ones) they do so *without addressing the distinct ethics of professional practices* as such (cf. Gunkel 2024, Kirk et al. 2024). In short, most of the scholarly discourse on AI moral alignment, in designated journals or handbooks devoted to AI ethics, do not tackle the challenge of professional, practice-specific ethics in AI.

Numerous policy and industry initiatives have followed suit, with applied ethical principles for AI agents published by multi-stakeholder organizations and universities, at both national and international levels (Ryan and Stahl 2021). The most recurring principles in these declarations, and public debates, tend to include beneficence, non-maleficence, autonomy, justice, and explicability, the latter being generally understood to encompass both intelligibility and accountability (Jedličková 2025). Anthropic have recently put forward the project of Constitutional AI, where the constitution framework incorporates established ethical guidelines, such as principles from the UN Universal Declaration of Human Rights (as well as rules drawing on non-Western belief systems, emphasizing collective responsibility, harmony, and respect for life), to ground its explicit goal of aligning the model with global ethics (Bai et al. 2022, Anthropic 2026).

From the "peripheral" perspective of specific professional practices, these ethical principles appear *overly general*. In fact, even their translation into practical guidelines aimed at policymakers and other stakeholders tends to remain highly broad in scope (Morley et al. 2023). Whether deontological or consequentialist or hybrid in their framing (cf. Krah and Tröschel 2026), the principles most commonly articulated and promoted - beneficence, non-maleficence, justice, fairness, among others - are assumed to be broadly applicable, treated as relevant benchmarks for most moral situations, regardless of the professional field they pertain too. However, suggesting adherence to "common good", "well-being", "non-bias", "remedy", "equity", "diversity", and "peace", among others (Ryan and Stahl 2021) is not, in itself, especially useful for the *distinctive roles and constraints of practice-specific ethics* within particular contexts and situations (cf. Wei and Zhou 2022).

Returning to the human context reminds us that many professional fields have developed their own ethical frameworks, designated for their professional practices. This is evident across diverse professions and practices and is distinct from the moral or political philosophy often evoked in the context of AI ethics. There are both applied and theoretical discourses *devoted to separate ethics* in medicine, education, law, software development, translation, and others (Freedman 1979, Peters 2015, Marino 2017, Wendel 2020, Gogoll et al. 2021, Koskinen and Pokorn 2021). As suggested by ethicist Anthony Pym in the case of translation (and one finds similar assertions in other professional fields): "any principles [of professional ethics] should ideally be involved in the activity itself and can in some way be drawn out of its discourses. This involves sketching a regional (i.e. non-universal) ethics, intended only for a particular set of social activities, and thus self-consciously unable to make grand pronouncements on any wider humanity" (2012, 4; italics in the original). Along these lines, Pym's professional ethics make a clear point of "not attempt[ing] to take a position on universal positions of right and wrong causes" (Pym 2012, 4). Similarly in medical ethics, it has been suggested that "the distinctive obligations of physicians derive from the role and function of medicine" (Rhodes 2019) – and comparable approaches can be found in other professional fields as well.

The reason for this is not only conceptual or ideological but also practical. Ethical discussions integral to professional practices are naturally richer and more attentive to the specificities of their *respective* domains than general or universal ethics. Due to the limitations of scope, this article uses examples mainly from the fields of medical, legal and translation ethics – but the same picture arises in other professions. In the case of translation ethics, for example, ethical approaches take into account the translator's commitment to the semantic properties of the source text and language, the communicative and functional needs of the commissioner of the translation, the functions of intercultural understanding for the stakeholders involved, among other considerations (Lambert 2023). In the case of medical ethics, considerations include the need to act for patients' benefit sometimes at the expense of others, to maintain their confidentiality, to ask intrusive questions, to intervene paternalistically in people's autonomy, among others (Rhodes 2019). The case of legal ethics also involves a narrower set of concerns and obligations than general morality, including some degree of partisanship for one's client, or striving for neutrality based on technical legal considerations, among others (Postema 1980). This should not give the impression of monolithic professional ethics. Rather, there is a largely pluralistic nature to the separate ethics literatures of professions and practices, as evident in the respective polemics in the fields of translation, legal practice, and bioethics (e.g., Koskinen and Pokorn 2021). In fact, this pluralism reflects well the particularities of each profession. Ultimately, the research literature demonstrates that the diverse values and considerations integral to each professional field, manifested in professional codes and debates, are more directly connected to the human stakeholders who depend on these practices in their everyday lives than any sweeping *general* system of ethics – even as the preference of one ethic over the other in each professional practice is *situation-dependent*, as well as a normative and *subjective* choice (Asscher 2025, 92-98).

This does not mean there is no need for general principles of morality in the conceptualization of decision-making in professional practices. Rather, debates in the field show that professional ethics have a separate existence from, yet are not independent of, general moral considerations (Gewirth 1986). There is no full consensus in the research literature on the *relationship* between the designated ethical considerations of professional practices and the broad considerations of general morality. But it is largely accepted that professional ethics *intersect* and *interact* with general ethics, and that situated ethical tensions are decided in the complex negotiation and consolidation between the two.

A prevalent view, found across the ethics literatures in fields such as law, education, medicine, and translation, argues that there is a *hierarchical relationship* between professional and general ethics, which should determine the interaction between them, and inform the ethical output in given cases. As put by Chesterman, "any professional ethic must be subservient to more general or universal ethics, since professions and practices only concern subsets of societies, just as societies are subsets of humankind as a whole, and humankind of organic life in general" (2001, 152). In the context of medical ethics, Benjamin Freedman specifically suggests that "Professional morality places professional values at a higher position in the ethical hierarchy" (1978, 10). From this perspective, the realm of professional ethics is separate from, but dependent on, general ethics, because the latter are generally subordinate to the former in cases of incongruence. Other views in the ethics literature suggest that professional ethics are not separate from general morality – but are rather grounded in, and ultimately should derive from, the same moral concerns that govern "ordinary" human interactions (Martin 1981, Koehn 2006). From this view, there should be no gap or distance that separates the claims of general and professional ethics, as professional ethics should effectively be shaped in the image of general ethics.

Without taking a normative stance on the matter ourselves, it is clear that some kind of *relationship* between professional and general ethics exists whenever decision-making involves a professional practice; and some equilibrium - often a compromise - is manifest in the tug of war between competing (general and practice-specific) ethical claims. It is this particular *point of balance*, realized in the *proportional distance* from the claims of either "side" of the ethics equation, which decides the ethical output (cf. Van Thiel and Van Delden 2010, 187-193).

Finally, it is worth mentioning the research literature on AI ethics which has focused on the political economy of professions in the neo-liberal era, and how it pertains to human professions and sustainability being threatened by AI (cf. Do Carmo 2025). Our theoretical goal of conceptualizing the practice-specific decision-making of AI systems is, in itself, orthogonal to questions about the effects of AI on the present and future of human professions whose practices are mentioned here. This is because, while we find LLMs to exhibit some form of ethical agency, the wider social implications of their appearance on the scene for the world of professions stem from institutional and state approaches to the *governance* of AI companies, and how AI development and deployment is *regulated* in today's political economy (Walter 2024). It is mainly this regulation (or lack thereof), together with common conceptions of AI in the public discourse - both of which are beyond the scope of this article – that will determine the ways in which AI affects the professional prospects and lives of translators, legal workers, programmers, etc. Conceptualizing the ethical underpinnings of practice-specific AI production should not imply the perspective, touted and oversimplified by AI companies and the media, that humans can be readily substituted in these professions by AI agents, and our article in no way endorses the unregulated pressure put on the respective professions. The opposite is true, particularly with regard to the moral and legal accountability provided by human professionals, and with regard to the need for long-term sustainability of human professions and industries (Tyson and Zysman 2022, Moorkens et al. 2024, Panezi and O'Shea 2025, Berz 2026).

This, however, does not mean we should evade the research in question. The increased lay use of AI for professional practices is a widespread reality, and as such requires academic analysis and description, whether we believe the political economy behind it is socially detrimental or not. Therefore, in the following section we conceptualize the increasingly autonomous AI decision-making in professional practices from an ethical perspective, without touching directly on the connection between public policy and regulation of AI companies, or the pervasive implications of AI in society. Still, it is hoped that the current article contributes indirectly to this discussion, not least by shining a light on the levels of complexity and subjectivity involved, which pertain to the immense moral responsibility that lies on the shoulders of the actors facilitating complex, competing ethical claims, not to be taken lightly or without the necessary expertise.

## 3. Ethics of professional practice in generative AI: four main dimensions

The main points entailed by the research literature serve as useful scaffolding for our efforts to conceptualize the ethical decision-making of generative AI in areas of professional practice. One can imagine such a framework of ethical agency as not categorically different from that of, say, a human who practices both translation and medicine: this person is expected to follow the ethics of translation when translating, and the ethics of medicine when practicing medical advice, while nevertheless subordinating these professional ethical frameworks to a higher set of general ethical principles, or otherwise negotiating between general and practice-specific ethics towards some nuanced compromise, when it is deemed necessary in light of the situational context in question.

It thus becomes clear that our conceptualization needs to involve several dimensions, or parameters, that *jointly* determine AI ethical agency in areas of professional practice. These dimensions are: 1) the different **professional ethics** embedded in AI systems; 2) the **general ethics** embedded in AI systems; 3) the hierarchical or other relationship between the **professional ethics** and **general ethics**; 4) the characteristics of **the situational context**. Let us first describe each of these four dimensions separately, before touching on how they interact, and then, in the next section, on how they pertain to our practical goals of empirical description and intervention.

### 3.1 Professional ethics

Professional ethics, or ethics of professional practice, are best represented by the divergent ethical approaches and polemics developed *within* a professional practice in consideration of applied cases. In AI models, just as we have seen in the research literature, professional ethics can be inherently *diverse*: there are no *unified* ethics of law, or translation, or medicine, or education, or coding, which are accepted by all. Moreover, existing ethical frameworks in professional practices are *historically-contingent* and never fixed over time (Backof and Martin 1991).

The ethics of professional practice are inherently *discursive*, in the sense that they are drawn from and are defined by the discourses of the practice, rather than being formally *rule-based* in the logical or syntactical sense. This is characteristic of how contemporary AI systems work: there are no set rules to decide on how to ethically approach a legal case, or a medical case, or a commissioned translation, etc., but rather an ethical repertoire of guidelines that tend to take *some narrative form* (Abdulhai et al. 2024, Zewail et al. 2026). The discursive essence of professional ethics may explain how varied principles in each professional discourse are learned, in some capacity, by foundational AI models in the pre-training phase. By this we mean that, regardless of the fact that internal decision-making processes remain largely opaque, the AI model reflects the data it trains on and exhibits some kind of professional ethics, as evinced by its production in practice-specific tasks. Of course, these ethics may appear to us partial, inconsistent or otherwise deficient, but that is for us to determine *empirically*, and perhaps reshape, if we wish to do so through post-training (as developed in the next section). Indeed, the empirical study of these phenomena is important because while LLMs' ethical approaches and guidelines are *drawn* from discourse, their effects in the world can be very tangible; they do not only feed back into the discourse but also affect human experiences and lives, in sometimes hazardous ways (Glickman and Sharot 2025).

Professional ethics in AI systems naturally vary and differ with reference to each professional practice. For the sake of simplicity, let us call Medical Ethics *ME*, Legal Ethics *LE*, and Translation Ethics *TE*. Within each of these respective fields, one tends to find divergent approaches and priorities (to enhance clarity, the use of acronyms in this section is limited only to *internally* differing instances). In Legal Ethics, for example, divergent ethical imperatives include loyalty to the client, fidelity to legal institutions and the rule of law, care and concern for human relationships, protection of third parties and the public good, among others (Rhodes 2019). In Translation Ethics, one finds potentially competing priorities such as faithfulness to the source text, adherence to the functional-communicative needs of the commissioner of the translation, long-term intercultural cooperation across cultures, accentuating cultural difference, among others (Koskinen and Pokorn 2021). And so on. Importantly, these practice-specific moral imperatives can be in direct contrast with each other in a given situational context: a fork-in-the-road conflict can occur, in which the implementation of one ethical stance comes at the expense of another, depending on the model's prioritization (Asscher et al. 2026).

In AI models just like in human practice, the implementation of ethical guidelines of professional practice – not only the broad umbrellas of *ME*, *LE, TE*, but also the divergent approaches and priorities within them – is *subject to interpretation*. An interpretative choice is ultimately needed to choose how exactly to apply an ethical strategy. There is no definitive answer as to what solutions best implement an ethical approach while prioritizing a certain goal; rather, there are divergent and often competing choices at any juncture and an interpretative lens is always needed to select one. For example, some ethical approaches to medical advice may require some form of paternalism towards a patient, but this paternalism can take very different forms (Sjöstrand et al. 2013). Certain ethical approaches to translation promote, above all else, a "faithfulness" to the semantic properties of the source text, but how this faithfulness is implemented, or which semantic properties the translation is most faithful to, can greatly vary depending on interpretation (Nida and Taber 1982). This inevitable subjectivity does not mean that different ethical frameworks are redundant. They differ in the guidelines that frame and direct the interpretation, which is still very important. Carrying out a professional practice in light of the (subjectively interpreted) ideal $LE_1$ will bring about different results from doing so considering the (subjectively interpreted) ideal of $LE_2$ or $LE_3$, and so on.

Finally, practice-specific professional ethics in AI systems **cannot be reduced to general or universal ethics, as they have unique imperatives that are attentive to the particularities of the profession** (Freedman 1978). They are distinguished from general ethics in that they may give different weight to the same values or have different pragmatic or moral considerations. The physician's commitment to patient confidentiality, the translator's obligation to the source text, or the solicitor's obligation to the local legal provisions in a particular jurisdiction, are cases in point.

### 3.2 General ethics

General ethics represent the second dimension in our framework. What stands out is their basic ***structural*** similarities with ethics of professional practice, from a practical AI perspective. Just like professional ethics, general ethics are **multiple** and **unfixed** – as evident in the field of political philosophy, and in the cultural dependency of value systems over time and space; notions of value pluralism have informed the discussion of AI ethics as well (e.g., Vallor 2016, Coeckelbergh 2022). General ethics are also **discursive**, in that they take narrative form and are represented in text, rather than being formally rule-based; and they are also open to **subjective interpretation**, as evident in how diversely they are carried out and perceived in applied ethics (Jobin et al. 2019). Their practical implementation, too, depends on the situational context. Importantly, the divergent and sometimes contrasting approaches within general ethics may belong to a Western or non-Western tradition, a conservative or liberal or progressive repertoire or any mix of the above, but that does *not* change any of the inherent structural properties described above.

Along these lines, and similarly to professional ethics, different principles in general ethics may somewhat overlap or lead to the same result, but they are distinct from each other in that they represent *a different hierarchy of priorities*, or, in the terms of political philosopher John Rawls, “lexical ordering” or “lexical priority” (Rawls 1988). For example, the superiority of basic liberties (freedom of speech, political participation, etc.) over greater social equality and economic distribution (in cases of direct conflict), proposed by Rawls, is one such hierarchy (Rawls 2017) – but there can be others as well.

We can therefore think of these general moral imperatives as $GE_1$, $GE_2$, $GE_3$…, and consider $GE_i$ as distinct from $GE_j$ if prioritization *i* may be in direct conflict with prioritization *j* in a given situational context (cf. Rao et al 2023; and Weidinger et al. 2023 for an implementation of Rawls' theory of justice in AI systems). This echoes what we have seen in professional ethics, where a fork-in-the-road conflict can occur, in which the implementation of one ethical stance comes at the expense of another (Crowder 2019).

For our purposes, it is worth emphasizing that, as a rule, **general ethics do not make an explicit claim over professional practices and fields** but rather apply to general ethics and morality in "ordinary" social life (Allen and Wallach 2009, Coeckelbergh 2020). That said, while general ethics have less of an applied nature inasmuch as professional practice is concerned, they are still applicable to human relations and life and are meant for implementation. As a result, there *is* a juncture between general ethics and professional ethics that allows for meaningful *interaction* between them. This interaction cannot always be resolved in easy conciliation. In each situational context, the interaction between general ethics and professional ethics can in fact embody deep tension as well. As noted by Benjamin Freedman, "Conflict between professional and ordinary morality does not arise, then, through happenstance but is an essential part of the description of the relationship between these two moralities" (1978, 10). That said, both ethical repertoires "speak the same language" inasmuch as they can have a "conversation" that effectively *resolves* the tension as an *equilibrium* is sought after and found.

For the sake of clarity, let us first formalize this tension - the potentially *conflicting* nature of ethical imperatives. Let us mark the differing approaches within Medical Ethics as $ME_1$, $ME_2$, $ME_3$, etc.; differing approaches in Translation Ethics as $TE_1$, $TE_2$, $TE_3$; in Legal Ethics as $LE_1$, $LE_2$, $LE_3$; and so on. And let us mark differing approaches in general ethics as $GE_1$, $GE_2$, $GE_3$, etc. It is true that, in a certain context, these different ethical approaches may have some pragmatic overlap, e.g., when a translation's fidelity to the source text ($TE_i$) *happens to align* with the general ethical goal of "diversity" ($GE_j$), and both can be satisfied in the same output. Or, when the medical imperative of keeping the patient's privacy ($ME_i$) *happens to coincide* with the general ethics of "common good" ($GE_j$), and both can be satisfied in the same output. However, this is not *always* the case; cases of tension and conflict can and do occur – for example, if the source text in the translation situation above is inciting and misogynistic; or if, in the medical situation above, keeping the patient's privacy may conflict with the safety of others. To conclude, general ethics and professional ethics do not *exclusively* impose, but can and may create a fork-in-the-road ethical conflict in which the implementation of one ethical stance *comes at the expense of another*.

In these fork-in-the-road cases, a direct conflict is *observable* in the output – it is evident, in practice, in how an ethical stance is implemented by an AI model or not (Liu et al. 2026), in line with the relationship between general ethics and professional ethics prescribed to it (more on this below). In the examples above, it would have been evident in how the text was translated, or, say, in a medical context, in whether the AI model prioritized respect for a patient's autonomy or the physician's duty to promote the patient's welfare when recommending how to respond to a patient's refusal of treatment. Another example would be a conflict that arises between the medical ethic of maximizing the effort to save an individual patient through a scarce treatment and the broader ethical principle of allocating limited resources fairly across the wider population. In this case, implementing one ethical stance can come at the expense of another, and the conflict is observable in the model's ultimate recommendation.

### 3.3 The situational context

Our framework's third dimension is the situational context, which provides the AI model with the concrete environment in which our abstract principles are *applied*, and the professional practice is actually performed. This happens when an AI system translates a text, administers medical advice, assesses or creates a legal document, codes a program, etc. The specific details may include, in a translation situation, the source text, the functional needs the commissioner expects the translation to achieve, the conventions in the target audience, among others. In a medical situation, the specific details may include, for example, the medical history of the patient and their parents, the communicative settings in which they share or don't share this history, the symptoms experienced, and severity of the described condition, the expectations and level of consciousness of the patient, among others. **The particularities of the context are therefore what gives rise to the tension between general ethics and professional ethics**, such that this tension exists, because they are the concrete backdrop for the negotiation and ultimate equilibrium established between *competing* ethical claims (more on this below).

Importantly, what the situational context consists of depends on the professional ethics involved, as they shine a light on what is seen to be pertinent. For example, in a medical situational context, the previous medical history of a patient's parents may count as relevant information. In a legal situational context, the set of legal provisions for having a right to work under a visitor visa may be relevant, if this concerns the professional practice at hand. It can also involve the stakeholders' specified needs, institutional requirements, or individual preferences. But the bottom line is that the situational context is what is seen to be crucial for the practice in question; **contextual information** **that is relevant to one profession is not necessarily relevant to another**. Of course, not all the contextual information is always available to the AI model; the more of it that is available, the fuller and more nuanced the practice-specific ethical decision-making can be.

### 3.4 The relationship between professional ethics and general ethics

Finally, our framework's fourth dimension is the prescribed relationship between professional ethics and general ethics. As we have seen in the research literature, the nature of this relationship is, in itself, a source of ongoing debate in the various professional fields and is rooted in both descriptive and normative work. This dimension is crucial because it defines the interaction between general ethics and professional ethics in the AI model, and how clashes between competing principles are to be resolved, in a given situational context (in whatever field – Medical Ethics, Translation Ethics, Legal Ethics, etc.). This principle applies regardless of if AI ethical agency is approached here from a deontological, virtue-based or consequentialist framework (cf. Tennant et al. 2025). These categorizations are less important for our purposes. What matters is that the situated interaction between these two ethical layers does not *always* have to impose an ethical conflict, but, as noted above, it can and often does. Operating under **both ethical layers simultaneously**, the AI model has to reconcile their competing demands, if they are conflicted, according to the rules of the relationship between them.

The relationship can be one of **subordination** in a certain direction: professional ethics can be perceived as fully subordinate to general ethics, so that whenever there is a conflict between them, the requirements of general ethics a-priori prevail. Under such a relationship, a lawyer would always have to subordinate their professional ethics – such as not representing clients with opposing interests - to the demands of general ethics. Or rather, if general ethics are perceived as subservient to professional ethics, the requirements of professional ethics are always implemented in the case of fork-in-the-road conflict, at the expense of general ethics (cf. Rhodes 2019). Otherwise, there can be a more **nuanced relationship between competing claims** that involves both hierarchy and, ultimately, compromise.

Inspired by political philosopher John Rawls' notion of **reflective equilibrium**, competing value-laden claims can be subjected to an iterative process of mutual adjustment, in which principles and considered judgments are revised until a sufficiently coherent position between them is reached (Rawls 1988). As recent works have shown, reflective equilibrium provides a useful conceptual analogy for thinking about the reconciliation of competing moral objectives in LLM training and fine-tuning (Anderson 2025, Ma et al.

2025). In machine learning, optimization similarly involves iteratively adjusting a model to improve its performance with respect to a specified objective - and when this objective involves multiple goals, constraints, or sources of feedback, it necessarily produces trade-offs between competing demands. The mechanism of reinforcement learning, for example, can optimize a reward function that combines different objectives, and results in the model's behavior reflecting the relative weighting assigned to them. This analogy should not, however, be taken to imply that optimization processes fully *constitute* reflective equilibrium: whereas mathematical optimization seeks a solution according to a formally specified objective, reflective equilibrium involves the substantive iterative revision and reconciliation of principles and judgments.

The relevance of Rawls' reflective equilibrium to how LLMs may resolve moral conflicts is also rooted in that it is a framework that seeks a compromise between *hierarchically* organized demands. Therefore, depending on the preferred professional ethics, general ethics, and the hierarchical relationship between them, the system adjusts outputs iteratively, until it reaches something like a reflective equilibrium between potentially hierarchical ethical layers (cf. Brophy 2026). Echoing Rawls' notion of "lexical priority", this is done by assuming moral principles' different orders of priority, where earlier principles must be fully satisfied before later ones apply, and the incoherence between hierarchically-ordered ethical priorities is **incrementally reduced** until reaching an equilibrium (Rawls 2017).

Here, it is worth noting that **some professional ethics are more inclined than others to have some intrinsic orientation to general ethics**, or to reverberate more universal dimensions in their practice-specific ethics. For example, some of the duties identified by Rhodes in her medical ethics are the respect for persons, concern for patients' welfare, and fairness, which are already recognizable as broadly human moral concerns – while others are much more specific to the medical practice (Rhodes 2019). Freedman similarly noted that "Medical confidentiality is a profession-specific norm, but when a patient poses a serious threat to another person, universal obligations to protect others can override it. This demonstrates that some professional ethics incorporate universal moral considerations more deeply than others, even while maintaining distinctive professional imperatives" (Freedman 1978). In Anthony Pym's translation ethics of long-term intercultural cooperation, despite being expressly "regional" and profession-specific, the flexible deontological principles encompass general long-term considerations of cooperation, including a bird's-eye view of translation's role in society, and how it may benefit a more heterogeneous and equal world – which corresponds well to some principles of general or universal ethics (Pym 2012). In practice, the degree to which $ME_i$ has a more universal inclination than $ME_j$, or $TE_i$ has a more universal inclination than *$TE_j$*, influences the iterative process by which the AI model may reach a reflective equilibrium with the conflicting general ethics. The incremental reduction of incoherence between a professional ethic and general ethic is directly shaped by these preliminary features: for example, a compromise sought between general ethics and *universally-oriented* professional ethics could be easier to reach than a compromise between general ethics and *regionally-oriented* practice-specific professional ethics.

## 4. Practical implications: how to *evaluate* equilibriums between general and professional ethics, and how to *reshape* them

Our conceptual framework has several practical and empirical implications. These are entailed by the fact that generative AI models are *already* performing professional practices on a massive scale, besides and alongside acting generally in the world as multi-task autonomous agents: they are *already* prioritizing certain ethics of professional practice over others and are *already* finding ways to reconcile general and professional ethics whenever a conflict arises between these potentially divergent ethical repertoires.

How do the major commercial LLMs acquire their professional ethics, which are then realized in user interactions? In what ways is the hierarchical (or other) relationship between general ethics and professional ethics instilled in these generative AI models, in the various professional fields to which they pertain? While we do not have access to the ways in which AI companies such as OpenAI, Anthropic, and Google, have trained their close-source models, there are several basic mechanisms by which major generative AI models can be assumed to have reached their current configurations and trade-offs between general and professional ethics. In the **pre-training stage**, a model is trained on immense corpora, which includes textual material pertaining to general ethics (human rights discourse, protocols of general morality, academic discourse on moral philosophy, etc.) as well as to professional practices (professional codes of ethics, professional forums on social media, actual practice-specific production, etc.). In this pre-training stage, the textual material that talks *about* the practice (e.g., critiques of translation, explanations about legislation) and the manifestations of the practice *itself* (e.g., translations, contracts) constitute data from which the model may learn associations between professional practices, norms, values, and patterns of appropriate behavior. Then, there is the **post-training stage**, where supervised fine-tuning through examples, and reinforcement learning through human feedback, adjust a model's performance and output in specific domains, including professional fields. These procedures may encode implicit ethical assumptions, insofar as the examples, evaluative criteria, reward models, or policies used to shape the model's desired behavior embody judgments about what constitutes an appropriate professional practice. Finally, the model's professional ethics can also be shaped by individual user preferences, which can be incorporated through **user interaction** into subsequent model improvement or training processes where providers permit and select such data. In this case, conversational memory, retrieval, and other personalization mechanisms condition a model's responses on information about a particular user without modifying the underlying model weights. These mechanisms therefore operate at different timescales: personalization can affect behavior within or across interactions, whereas incorporating user interactions into subsequent training can influence later model versions.

With these mechanisms in mind, there are two important practical implementations that are entailed by our conceptual framework. These are: 1) offering an in-depth ***evaluation*** of the current ethical preferences embedded in major LLMs; and 2) suggesting how to ***intervene*** in these preferences in order to change them. As we have already noted above, there are no unequivocally "correct" answers, not in the prioritization of ethical approaches nor in how they are interpreted and implemented. Therefore, both the ethical assessment and ethical intervention would have to be rooted in some degree of interpretation and subjective preference on the part of the user.

The first of our two practical goals, then, is the **evaluation** of the models' ethical preferences when they perform professional practices. Whether our evaluation is qualitative or quantitative in nature, it should be rooted in the conceptual dimensions of the relationship between AI systems' general and professional ethics outlined in this paper. This means the need to consider, in a series of multiple situational contexts: 1) which of the ethics of professional practice is prioritized; 2) which of the orientations of general ethics is prioritized; 3) What is the hierarchical relationship realized between 1) and 2). Multiple situational contexts are necessary for the evaluation to identify the deontological threshold at which the equilibrium rests. Imagine, for example, a case where confidentiality between an AI model, acting as a legal advisor, and their user/client is found to be breached when the confidential information pertains to an imminent threat to another person; and another case where the confidentiality is already breached under a weaker condition, e.g., when the confidential information pertains to a threat that is not imminent but rather general and contingent on other actors. The equilibrium between general and professional ethics is different in these two cases, and our framework of evaluation should be able to identify this difference in a model's behavior. The evaluation would be able to capture the model's professional ethics only if it examines the model's ethical behavior in a series of similar but not identical scenarios (cf. Jin et al. 2018).

Most evaluations of human ethical pluralism in the context of the tensions between general and professional ethics have been ***qualitative***, and involve an interpretation of ethical phenomena, carried out with the main epistemologies of the human sciences. These often, though not always, involve taking a *normative* stance and advocating for a certain relationship between the two ethical repertoires (cf. Freedman 1978, Wendel 2000, Chesterman 2001, Pym 2012, Rhodes 2019). A qualitative evaluation of AI professional ethics can be similarly fruitful, especially if it includes self-reflective modes of interpretation, acknowledging the researcher's own position and taking it into account in the analysis.

For a ***quantitative*** evaluation, one would need a benchmark against which the existing ethical equilibrium could be compared. The evaluation could then quantify the divergence between this benchmark and the model's existing configuration between its general ethics ($GE_i$) and professional ethics ($PE_j$) in a particular context, for example using Kullback-Leibler divergence (cf. Go et al. 2023; for the sake of clarity, the use of the *GE* and *PE* acronyms has been limited to the quantitative formulation in this section. $GE_i$ and $PE_j$ are used to designate a *particular* orientation of general and professional ethics throughout).

Given a situational context, then, the benchmark would provide a quantitative representation of our three main criteria: (1) the model's underlying $PE_j$; (2) its underlying $GE_i$; and (3) the relationship between the two. The model's ethical decision-making could then be represented as a distribution of probabilities over a spectrum of its possible judgments or actions, allowing the divergence between the benchmark and the model's $GE_i$-$PE_j$ equilibrium to be systematically quantified. As noted above, this approach requires evaluating the model across a *series* of similar but non-identical scenarios, rather than treating its response to a single scenario as representative of its ethical equilibrium.

The quantitative nature of this representation should not obscure the fact that our benchmark is ultimately subjectively decided by the evaluator. The benchmark can stand for the consensual view in the field (e.g., the largely accepted view in medical debates that *PE* are superior to *GE* when it comes to confidentiality); but it can also stand for a subversive challenge to the prevalent conventions, by suggesting as a yardstick a less common *GE-PE* equilibrium (e.g., when postcolonial translation scholars advocate for appropriating the translation of the source text in order to undermine social inequalities between ethnic groups, even at the expense of professional expectations of semantic faithfulness); or it could stand for any other approach on the $GE_i$-$PE_j$ spectrum. Whatever the case, our evaluation requires the quantitative representation of a benchmark, specified by the evaluator, in order to measure the distance of AI models' ethical preferences from this benchmark (cf. Awad et al. 2018, Haas et al. 2026). Our proposed form of ethical evaluation therefore consists of both a posited benchmark and the measured distance from it.

The second practical aspect our article entails is that of **intervention**, namely, trying to reshape the model's $GE_i$-$PE_j$ relationship to some extent. In a sense, this is a natural continuation of the quantitative evaluation described above, because the intervention - the attempt to "push" the $GE_i$-$PE_j$ relationship in a particular trajectory - can be usefully formulated as reducing the divergence between the model's existing $GE_i$-$PE_j$ relationship and a normatively specified $GE_i$-$PE_j$ relationship.

To demonstrate this with one of our previous examples, let us suppose we want to intervene in the professional ethics of our AI model acting as a legal advisor, as it is manifested in its $GE_i$-$PE_j$ equilibrium regarding the level of professional confidentiality it keeps in the case of the potential of a public threat. Assume we have evaluated the existing $GE_i$-$PE_j$ relationship by examining the model's ethical behavior in a series of similar but not identical scenarios, where the level of threat to other people, as revealed in the confidential information shared by the user with the model, increases incrementally. We have pinpointed the existing equilibrium and want to slightly push it towards the $GE_i$ ('prevent harm') end of the spectrum, so that the model will breach the professional confidence with the user requesting legal advice already under a *weaker condition* of threat level. To that end, we could use a reward function that brings the model closer to our desired professional ethics, which in this case relates to thresholds of legal confidentiality.

In formal mathematical terms, this reward function can be described as follows:

$$R_{total}=\alpha R_{prof}-\beta D_{KL}(\pi \| \pi_{ref})$$

where $R_{prof}$ rewards alignment with the desired professional ethics, while the KL-divergence term $D_{KL}(\pi \| \pi_{ref})$ penalizes deviations from the model's existing policy $\pi_{ref}$. The relative values of $\alpha$ and $\beta$ determine the trade-off between improving the model's alignment with the desired professional ethic and preserving the model's existing behaviour. The definition therefore does two separate things simultaneously:

1. It incentivizes alignment with the desired professional ethics (e.g., the confidentiality threshold we consider appropriate) through $R_{prof}$;
2. It constraints the extent to which the model departs from its existing policy through $D_{KL}(\pi \| \pi_{ref})$.

The optimizing function finds the best compromise between these two objectives. In less formal terms, what we are doing is refining a **deontological threshold** - pinpointing exactly where the threshold will be for prioritizing one ethical imperative over another - by adjusting these competing claims until they align. Concretely speaking, the desired ethical orientation can be promoted through interventions at different stages and levels of the AI system. These could include introducing relevant textual material into the training data during pre-training, applying the mechanisms described above during post-training, or using interactive machine learning and personal corpora to adapt the model through subsequent interaction. Alternatively, interventions could occur at inference time, for example through prompting, retrieval, external constraints, or other forms of behavioral steering, without modifying the underlying model parameters. Finally, the environment of the human-computer interaction itself could be modified (for instance, through institutional rules, feedback mechanisms, or the configuration of other agents) in order to incentivize the emergence of a desired $GE_i$-$PE_j$ relationship. In technical terms, within each of these different approaches and stages, interventions can be implemented through bottom-up, top-down, or hybrid mechanisms, in line with recent directions in the literature (cf. Tennant et al., 2025).

## 5. Conclusion

In recent years, AI systems have been manifesting professional ethics on the ground, as lay users and autonomous AI agents increasingly act in professional domains by initiating practices such as legal and medical advice, translation, coding, and more. By virtue of their production, today's LLMs effectively implement one form or other of professional ethics, with ethical consequences for all stakeholders involved. Experienced human professionals may have a higher proficiency in ethical reflection and reasoning than LLMs, but, for reasons mentioned in the first two sections, the volume of LLMs' ethically implicated decision-making in professional practices is likely to continue expanding. Against this backdrop, the current article has conceptualized the various dimensions of professional ethics in generative AI, and suggested how to empirically evaluate the phenomenon, and, if deemed necessary, how to intervene in it.

Perhaps most importantly, our conceptualization put the spotlight on AI models' professional ethics by emphasizing their separate existence from, and prescribed relationship with, general ethics, and on the mechanisms through which they reach an equilibrium in situational contexts that involve ethical conflict. It is through this equilibrium, we have suggested, that certain professional ethics are prioritized and implemented in practice - regardless of whether these ethics take the form of a deontological, consequentialist, or hybrid approach. We have also emphasized the dimension of subjectivity involved in the evaluation and benchmarking of AI models' professional ethics, though this subjectivity should not imply that the current article endorses a *relativistic* view of either professional or general ethics - as if all ethical options are equally valid or should be considered as such. The question of moral relativism belongs to a normative debate that is beyond the scope of this article. We have considered here only the *structural* features of the relationship between professional ethics and general ethics, and their implications for systematic evaluation and intervention.

The disruption caused by AI to the world of professions – translators, lawyers, healthcare providers, among others – requires efforts of governance and regulation in order to achieve sustainability for these crucial human industries. This should be considered an important ethical goal for academia as well, notably by supporting the academic departments and research areas relevant to these fields. In addition, there is a growing need to describe and evaluate the quickly expanding situation on the ground, where LLMs are performing ethically-implicated professional practices on a grand scale. Such empirical evaluation of professional ethics in generative AI will advance us toward a deeper and broader understanding of an increasingly AI-infused world of practice.